\documentstyle[11pt,moriond,epsfig]{article}

\def\Journal#1#2#3#4{{#1} {\bf #2}, #3 (#4)}

\def\mnras{\em MNRAS}
\def\apj{\em ApJ}
\def\aa{\em A\&A}
\def\aj{\em AJ}

\def\be{\begin{equation}}
\def\ee{\end{equation}}
\def\bea{\begin{eqnarray}}
\def\eea{\end{eqnarray}}

\begin{document}
\vspace*{4cm}
\title{ARCS STATISTICS AS A PROBE OF GALAXY EVOLUTION}

\author{J. BEZECOURT}

\address{Observatoire Midi--Pyr\'en\'ees, 14 avenue E. Belin,\\ 31400
Toulouse, France}

\maketitle\abstracts{Number counts and redshift distribution of gravitational
arcs are computed in the field of massive clusters of galaxies 
to probe the universe at high redshift.
Using an accurate modelling for the cluster mass distribution
and a model for the spectrophotometric evolution of galaxies, the
redshift
distribution of gravitational arclets is computed in the field of
cluster
Abell 2218 and in the Hubble Deep Field where a cluster is artificially
located. Counts are very well reproduced in the $B$ band but
an important population appears at high redshift which is not seen in
deep spectroscopic surveys. Unfortunately, the very high 
sensitivity of the counts with respect to the model for galaxy evolution 
and to the mass distribution prevents from estimating the cosmological
parameters with arcs statistics. Future works have to concentrate on high
redshift clusters and take advantage of objects with smaller distortions.}

\section{Number counts and redshift distribution of gravitational
arclets}
A powerful way to investigate the population of high redshift galaxies 
is to use the
magnification by the gravitational potentiel of massive clusters of
galaxies. The spectrophotometric evolution of galaxies can then be
probed by 
counts and redshift distribution of gravitational arcs.

The number counts of gravitational arclets in the field of a massive
cluster of galaxies are a competition between,
first, the magnification of the luminosity by the cluster
potential that makes more objects visible and, second, the surface 
dilution that
decreases the surface density of arclets by the same factor as for the
magnification.
If the surface density of galaxies up to magnitude $m$ is 
$n(<m)=n_0 10^{\alpha m}$, 
the ratio of the density of arcs over the density of field galaxies is 
${n_{arc}(<m)\over n(<m)}= M^{2.5 \alpha -1}$
where $M$ is the magnification.
For observations at faint level in the blue band, the counts remain roughly
unchanged while from the $R$ to $K$ wavebands ($\alpha <0.4$) a depletion 
takes place with respect to an empty field (Broadhurst 1995, 
Fort et al. 1996).

The number of arcs brighter than magnitude $m$ with an axis ratio
greater than $q_{min}$ and a surface brigthness brighter than $\mu_0$
within the field of a cluster of galaxies is:
$$N(m,q_{min},\mu_0) = \sum_{i} \int_{z_l}^{z_{max}} \int_{q_{min}}^{\infty}
S(q,z) \, \int_{L_{min}}^{L_{max}} \Phi_i(L,z) \, dL \, dq \,
{dV \over dz} \, dz $$
The sum is over the different morphological types $i$. 
$z_l$ is the lens redshift and $z_{max}(\mu_0,i)$ is the redshift cutoff
corresponding to the limit in central surface brightness $\mu_0$.
$S(q,z, H_0, \Omega)$ is the angular area in the source plane
(at redshift $z$) that gives arcs with an axis ratio between $q$ and
$q+dq$.
The luminosity $L$ is given by the model of Bruzual and
Charlot (1993, GISSEL96) using the results of Pozzetti et al. (1996).

It was required as
a preliminary step that counts and redshift distribution in empty field and
in various wavebands are correctly reproduced.
We used the model for galaxy evolution of Bruzual and Charlot (1993)
with the prescriptions of Pozzetti et al. (1996) (see B\'ezecourt et al. 
1998 for details).
In order to check the sensitivity to the cluster mass distribution,
three different potentials are used for cluster A2218.
The simpler one is a singular isothermal sphere (SIS) whose velocity
dispersion is determined by the location of a giant arc at $z=0.702$
($\sigma=1031\,km\,s^{-1}$).
The second one is a bimodal mass distribution centered on the two main
galaxies of the cluster (Kneib et al. 1995).
The more complex potential used was obtained by Kneib et al. (1996)
and includes galaxy scale components to account for the numerous
substructures in the cluster. 
Unless specified otherwise, $H_0=50\, km\, s^{-1}\, Mpc^{-1}$, $q_0=0.5$
and $\Omega_{\Lambda}=0$.

\section{Results}
\subsection{Absolute number counts in cluster A2218}
The detection of elongated objects in A2218
was performed in the frame of the WFPC2 HST image
(6200 sec.) in filter F702W (Kneib et al. 1996).

\begin{figure}
\centerline{
\psfig{figure=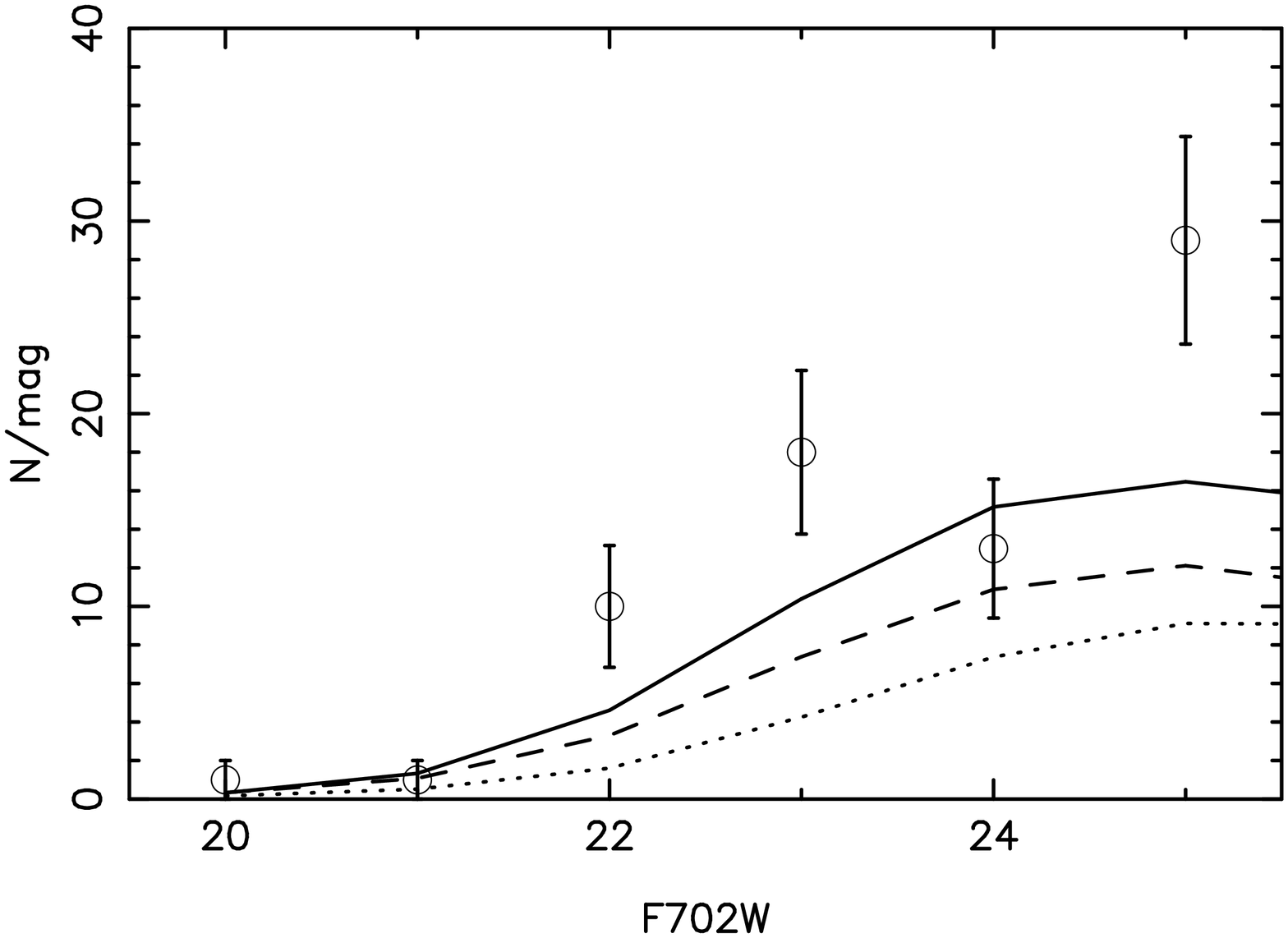,width=5cm,angle=-90}
\hskip 0.5cm
\psfig{figure=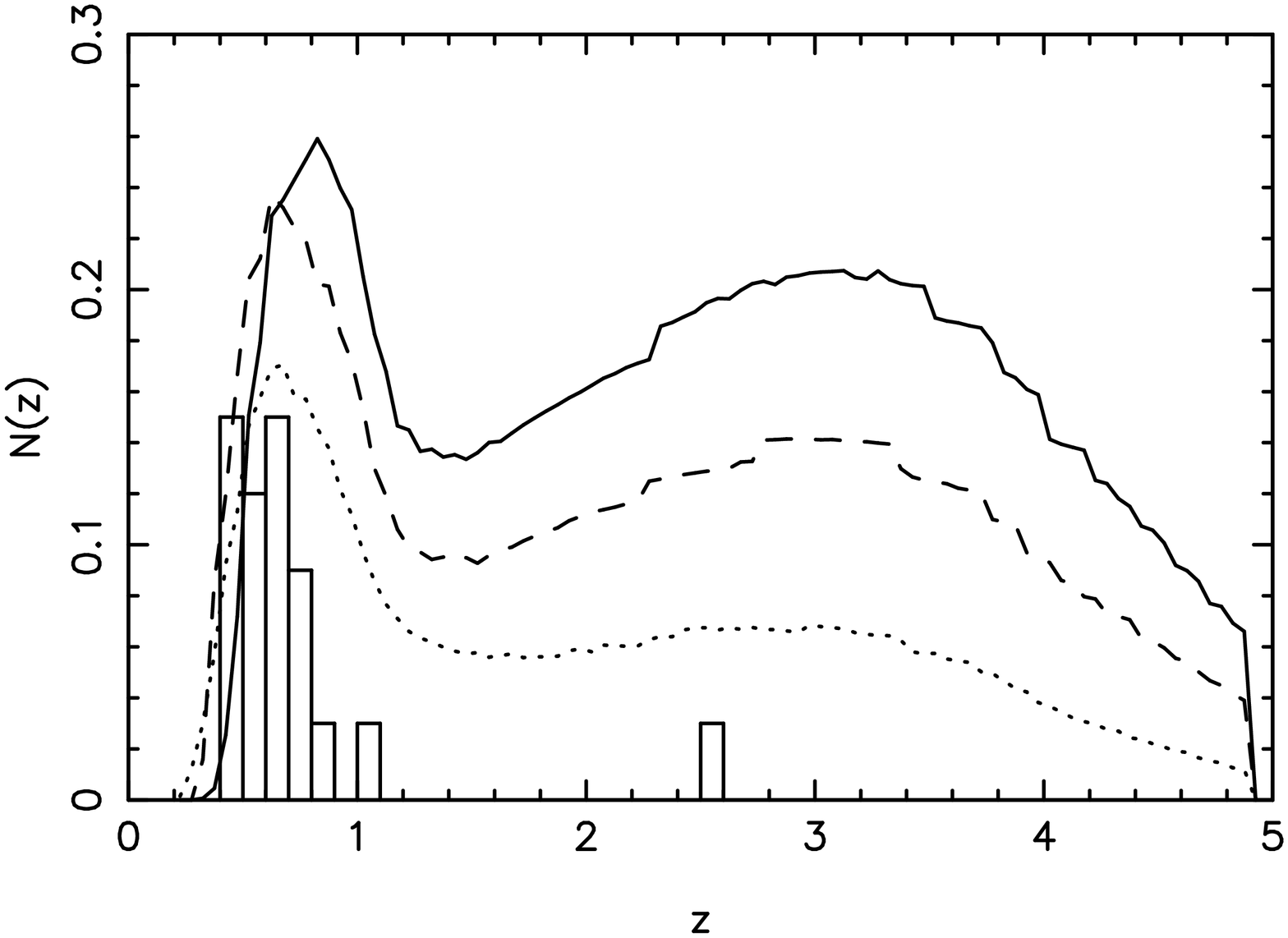,width=5cm,angle=-90}
}
\caption{Left: Number counts of arclets in A2218 with the F702W filter,
per bin of one magnitude ($a/b>2$, $\mu^0_{F702W}\leq 25.5$).
$\circ$: observed counts. Predicted counts are displayed for three
mass distributions: SIS (dotted line), bimodal mass distribution of Kneib
et al. (1995) (dashed line) and a mass distribution including
substructures from Kneib et al. (1996) (solid line).  
Error bars correspond to statistical fluctuations.
Right: Redshift distribution of arclets in A2218 per bin of 0.05 in
$z$ ($a/b>2$, $R_{F702W}\leq 23.5$ and $\mu_R \leq 24$). Same notations as for
the left figure. The histogram corresponds to the redshifts compiled by 
Ebbels et al. (1997), ordinate is in arbitrary units. 
}
\label{fig-2218}
\end{figure}

The observed and predicted number counts of arclets are in very good
agreement in the $B$ band considering the mass distribution of Kneib et
al. (1996). Counts in the $R$ band
are shown in Figure \ref{fig-2218}.
The predicted total number of arcs ($R_{F702W} \leq 23.5$ and $a/b\geq 2$)
with the best mass model is lower than the observed number
by a factor of 1.8.
The bimodal model of Kneib et al. (1995) and the SIS model underpredict the 
counts by  
factors of respectively 1.5 and 2 with respect to the model of Kneib et al. 
(1996) including
galaxy scale potentials. This is a clear justification for the use of
accurate mass distributions in statistics on gravitational arcs.

\subsection{Redshift distribution}

The redshift distribution of arclets selected in the F702W filter is presented 
in Figure \ref{fig-2218}
and can be partially compared to the results of a spectroscopic
survey of arclets (Ebbels et al. 1997). 
However, this survey may be biased in the arclets selection
criterion and in the ability to determine a redshift which mostly
depends on the existence of emission lines in the optical part of the
spectrum.
This observed distribution peaks at a value of
$z\simeq 0.6$, with only two objects at $z>1$. Figure \ref{fig-2218}
shows the redshift distribution of arclets expected with the same
selection conditions than those adopted by Ebbels et al. (1997).
A peak also appears at a similar redshift ($z\simeq 0.8$, solid curve). The 
slight difference between the observed and the predicted peak is due to the
hypothesis of circularity for the sources which prevents from obtaining
arcs at low redshift just behind the cluster because of a less efficient
lensing power. 

However, another population of objects is expected at $z>2$
which is not seen in the data. This high
redshift tail is mainly produced by elliptical-type galaxies and
shows the limit of the evolution model for young galaxies and early
epochs.

\section{Cosmological parameters}
\begin{figure}
\centerline{
\psfig{figure=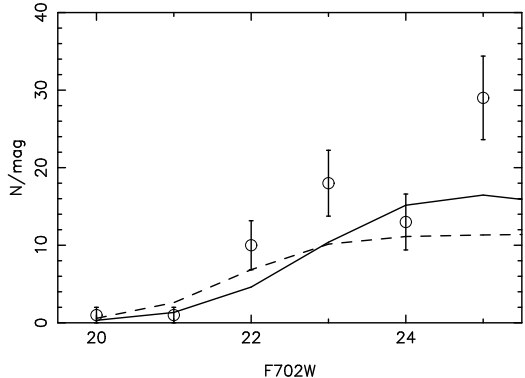,width=5cm,angle=-90}
}
\caption{Same figure as Figure \ref{fig-2218} 
for $q_0=0.5$ (solid line) and $q_0=0$ (dashed line)
with the complex mass distribution of Kneib et al. (1996).}
\label{fig-q0}
\end{figure}

Figure \ref{fig-q0} shows the predicted counts in A2218 for two
different values of $q_0$, 0 and 0.5. The difference between the two
curves is of the same order of magnitude as the uncertainty due to
the mass distribution between mass modellings of Kneib et al. (1995) and
(1996). This means that $q_0$ cannot be inferred from counts of
gravitational arcs. A low $q_0$ doesn't help to reconcile with
the data, hence one could argue that it implies a high value for 
$\Omega_{\Lambda}$
as it would increase the counts. However, this is unlikely as the
counts of arclets in the $B$ band are very well reproduced and a similar 
analysis
performed in cluster A370 (B\'ezecourt et al. in preparation) doesn't
require a high value of $\Omega_{\Lambda}$.
Moreover, uncertainties in galaxy evolution are too high to allow for
a determination of $\Omega_0$ and $\Omega_{\Lambda}$ with arcs counts.

\section{The Hubble Deep Field}
The results presented above can suffer from a privileged line of sight
in the direction of A2218
and a representative view of the distant universe is then needed. This sample
of distant galaxies is given by the images of the Hubble Deep Field 
(Williams et al. 1996). The model developped in the
previous sections can be pushed to fainter limits by locating artificially 
the mass distribution of cluster A2218 (Kneib et al. 1996) in front of the HDF.
Miralles and Pell\'o (1998) have determined photometric redshifts for a 
sample of 1400 galaxies in the HDF enabling to lens these objects by 
a cluster like A2218. 

\begin{figure}
\centerline{
\psfig{figure=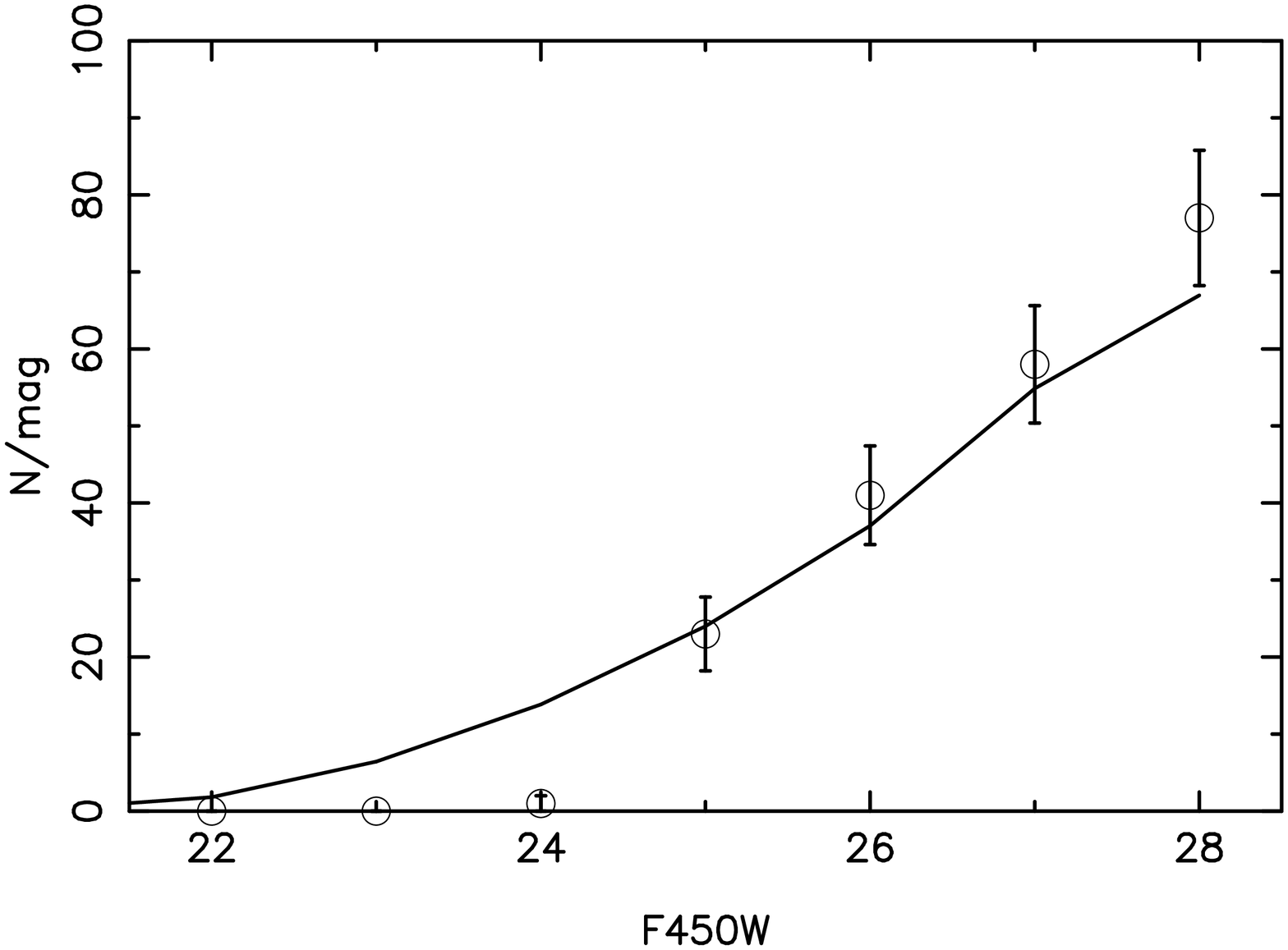,width=5cm,angle=-90}
\hskip 0.5cm
\psfig{figure=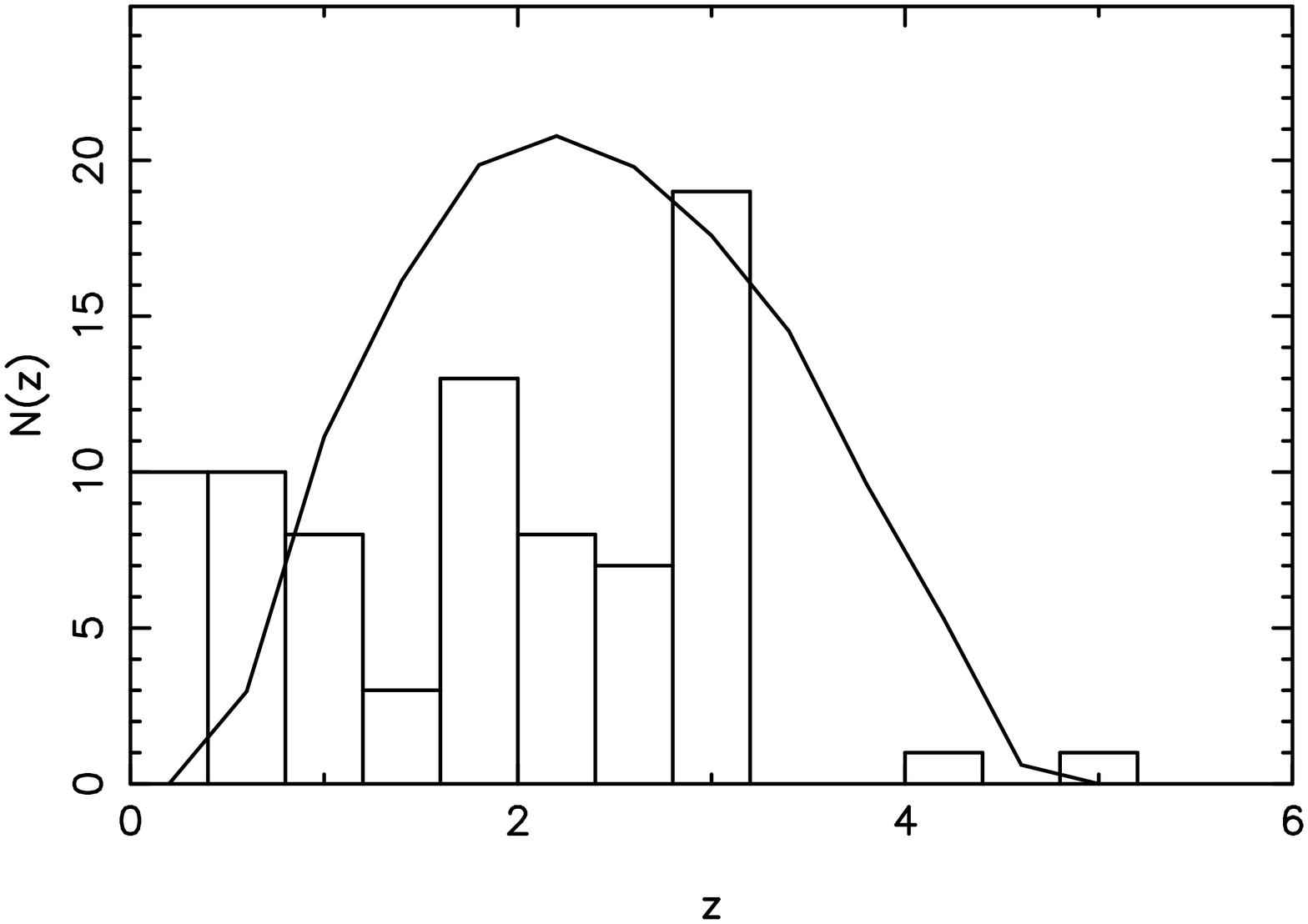,width=5cm,angle=-90}
}
\caption{Left: Counts of arclets in the HDF lensed
by A2218 artificially located in front of it (arclets with 
an axis ratio $a/b>2$).  $\circ$: simulated ``observed" counts. 
Solid line: predicted counts for 
the lens model by Kneib et al. (1996) and $q_0 =0.5$.
Right: Redshift distribution of arclets in the HDF per bin of 0.4 in
$z$ for $a/b >2$ and $B_{F450W}<27.5$ for the
simulated ``observed" arclets (histogram) and the predicted
distribution (solid line).}
\label{fig-hdf}
\end{figure}

The predicted counts of arcs in the field of the HDF (F450W filter)
with an axis ratio greater than 2 
are in excellent agreement with the counts of simulated arcs
(Figure \ref{fig-hdf}).
The redshift distribution resulting from the combination of the mass model 
with the evolution model for objects selected in the $B$ band
shows a deficit at $z<0.5$ due to the circularity of sources. On the contrary, an
excess of sources at high-$z$ appears as in Figure \ref{fig-2218}.
A different approach is then needed for
galaxy evolution at early epochs, particularly in the hierarchical scenario.

\vskip -0.1cm
\section{Conclusions}
It is now obvious that arcs statistics can only be performed with 
mass distributions including substructures, too simple potentials should
be avoided.
Number counts of arclets are in good agreement with observations in the
$B$ band and the redshift distribution of arclets at $z \le 1.0 $ 
is correctly predicted by the model. The important population
of arclets expected at $z \ge 1.0 $, which is not observed in
spectroscopic
surveys, is highly dependent on the modelling of high redshift
ellipticals and the role of dust absorption in the rest frame UV.

Unfortunately, the geometry of the universe cannot be determined with
arcs statistics because of
too important uncertainties in clusters mass distributions and in the model for
galaxy evolution.
Geometrical effects have now to be investigated with 
clusters at redshift greater than 0.5 considering also the weak 
lensing regime. High-$z$ lens efficiency 
relates more directly to cosmology and counts of all the background objects 
in a cluster field enable accurate measurement of 
the magnification bias that probes the high redshift population.

\vskip -0.1cm
\section*{Acknowledgments}
\vskip -0.1cm
I wish to thank the group of cosmology of the Observatoire
Midi--Pyr\'en\'ees of Toulouse for very fruitful discussions on galaxy
evolution and gravitational lensing.
I'm also grateful to Yannick Mellier and to the European Union 
for financial support.

\end{document}